\documentclass[%
 reprint,
superscriptaddress,
 amsmath,amssymb,
 aps,
 prl,
]{revtex4-1}
\usepackage{graphicx}
\usepackage{dcolumn}
\usepackage{bm}
\usepackage{natbib}
\usepackage{hyperref}
\usepackage{siunitx}
\usepackage[caption=false]{subfig}
\usepackage{booktabs}
\usepackage[export]{adjustbox}
\usepackage{multirow}
\usepackage{physics}
\usepackage{xcolor}
\usepackage{accents}

\begin{document}

\title{High-speed and high-connectivity two-qubit gates in long chains of trapped ions}

\def\ANU{Department of Quantum Science and Technology, The Australian National University, Canberra, ACT 2601, Australia}

\def\IonQ{IonQ, Inc., College Park, MD, USA}

\author{Isabelle Savill-Brown}
\email{Isabelle.Savill-Brown@anu.edu.au}%
 \affiliation{\ANU}
  \author{Joseph J. Hope}
 \affiliation{\ANU}%
    \author{Alexander K. Ratcliffe}
 \affiliation{\IonQ}%
  \author{Varun D. Vaidya}
 \affiliation{\IonQ}%
    \author{Haonan Liu}
 \affiliation{\IonQ}%
  \author{Simon A. Haine}
 \affiliation{\ANU}%
   \author{C. Ricardo Viteri}
 \affiliation{\IonQ}%
\author{Zain Mehdi}
 \email{Zain.Mehdi@anu.edu.au}%
 \affiliation{\ANU}%

\date{\today}

\begin{abstract}
We present a theoretical study of fast all-to-all entangling gates in trapped-ion quantum processors, based on impulsive excitation of spin-dependent motion with broadband laser pulses. Previous studies have shown that such fast gate schemes are highly scalable and naturally performant outside the Lamb-Dicke regime, however are limited to nearest-neighbour operations. Here we demonstrate that impulsive spin-dependent excitation can be used to perform high-fidelity non-local entangling operations in quasi-uniform chains of up to $40$ ions. We identify a regime of phonon-mediated entanglement between arbitrary pairs of ions in the chain, where any two pairs of ions in the chain can be entangled in approximately $1.3-2$ centre-of-mass oscillation periods. We assess the experimental feasibility of the proposed gate schemes, which reveals pulse error requirements that are weakly dependent on the length of the ion chain and the distance between the target qubits.
These results suggest entangling gates based on impulsive spin-dependent excitation presents new possibilities for large-scale computation in near-term ion-trap devices.
\end{abstract}


\maketitle

Trapped ion crystals are a leading platform for scalable quantum information processing~\cite{Bruzewicz2019b}, owing to the exceptional stability of atomic qubits~\cite{Langer2005,Harty2014,Wang2021a} and their exquisite control with electromagnetic and focused optical fields~\cite{Wright2019,Cetina2022,Chen2024}. Trapped-ion processors have demonstrated the highest fidelity single qubit and multi-qubit quantum logic operations~\cite{Ballance2016,Ryan-Anderson2024,Moses2023,Leu_2023,Cai2023b}, and have further been used for demonstrations of blind and distributed quantum computing~\cite{Drmota2024,Main2024}, proof-of-principle quantum algorithms~\cite{Debnath2016,Hempel2018a,Wright2019,Nam2020,Liu2025,Mayer2024}, and quantum error correction~\cite{Zhang2020c,Ryan-Anderson2021a,Egan2021,Nguyen2021,Postler2024,Mayer2024,Paetznick2024}.

Ion-trap quantum processors feature high connectivity between spatially-separated qubits, mediated by common vibrations of ions within a single trap~\cite{Wright2019,Debnath2016,Figgatt2019,Hou2024,Grzesiak2020,Chen2024} or by shuttling ions between trapping zones~\cite{Kielpinski2002,Wan2019,Kaushal2020a,Pino2021,Moses2023,Mordini2025,Akhtar2023}. This offers a significant reduction in resource overhead to connect distant qubits compared to platforms with limited connectivity, such as neutral atoms~\cite{Graham2022} and superconducting circuits~\cite{Arute2019}, and has further advantages for error-corrected computation using low-density-parity-check codes~\cite{Ye2025a,Tremblay2022,Bravyi2024}. However, quantum logic operations in trapped-ion devices can be slower than other platforms~\cite{Cai2023b,Linke2017}, limiting absolute computation speeds and the number of operations that can be leveraged against the long coherence times of trapped-ion qubits. This work addresses this challenge in trapped-ion quantum processing.

One pathway to speeding up quantum logic rates in trapped-ion systems is to employ `fast gate' protocols~\cite{Garcia-Ripoll2003,Duan2004a,Bentley2013,Gale2020a,Ratcliffe2020,Wong-Campos2017a}, where ions are subject to high-intensity broadband light fields resonant with the qubit energy separation that impulsively (rather than spectroscopically) excite qubit-state-dependent motion of the ion crystal. Theoretical studies suggest fast gates can enable MHz quantum logic rates without reduction of gate speed in scaled ion crystals~\cite{Taylor2017,Ratcliffe2018,Mehdi2020b:2D,Mehdi2021e}. In addition to challenging laser control requirements, a critical limitation of existing fast gate protocols is the limitation to nearest-neighbour operations~\cite{Bentley2013,Taylor2017,Mehdi2021}, which loses the significant advantage of `all-to-all connectivity' of conventional trapped-ion quantum processors. 

In this Letter, we present a class of highly-connected fast two-qubit (2Q) gates in long trapped ion chains controlled by sequences of spin-dependent kicks (SDKs) driven by broadband laser pulses. We identify two regimes of fast gate solutions: \emph{supersonic gates}, which are faster than the speed of sound in the trap; and \emph{subsonic gates}, which are mediated by collective vibrations of the ion crystal. The latter has a fundamentally non-local character which enables 2Q gates between distant qubits in chains containing tens of ions. Through simple numerical experiments, we extract a threshold 2Q gate speed of approximately $1.3$ centre-of-mass oscillations of the trapped-ion crystal between regimes. This threshold, which we suggest is a fundamental speed limit for phonon-mediated quantum logic in trapped-ion systems, separates the subsonic and supersonic fast gate regimes. We show that fast gates in the subsonic regime can be applied to nonlocal qubit pairs in linear chains of up to forty ions, enabling `all-to-all' connectivity with quantum logic rates in the range \SI{100}{\kilo\hertz} - \SI{1}{\mega\hertz}. Furthermore, we find the pulse requirements do not increase with the number of qubits for chains containing tens of ions, paving the way for high-speed quantum logic in ion-trap processors with dozens of physical qubits.

This work is supported by a companion manuscript~\cite{CP}, which elaborates on the theoretical design, parameter dependence, and robustness of fast 2Q gates in long ion chains.

\emph{Gate connectivity in trapped-ion chains.---} Conventional 2Q gate mechanisms are based on geometric phase gates using qubit-state-dependent driving of individual motional modes of the ion crystal on adiabatic timescales~\cite{Sorensen2000,Duan2001,Lee2005,Ballance2016}. The key advantage is that phonon modes are global properties of the crystal, which enables nonlocal quantum logic operations without transport of the physical ion qubit. Modern modifications to quasi-adiabatic mechanisms enable gate durations approaching the motional timescale of the crystal~\cite{Garcia-Ripoll2005, Steane2014, Palmero2017, saner2023breaking}, though this has only been demonstrated in two ion systems~\cite{saner2023breaking, Sch&xe4;fer}. For chains containing tens of qubits, controlling off-resonant excitation of spectator modes becomes increasingly challenging~\cite{Leung2018,Bentley2020,Cheng2023a,Landsman2019,Liu2023, zhuTrappedIonQuantum2006}, resulting in slowing gate speeds approaching $1$ms (compared to microsecond motional timescales)~\cite{Debnath2016,Figgatt2019,Grzesiak2020a,Chen2024}. The increased gate duration results in greater sensitivity to trap heating~\cite{Ballance2016}, further limiting the scalability of these protocols. 

The non-adiabatic nature of the fast gate mechanism enables quantum logic operations faster than the ions' motion, which suppresses crosstalk in scaled crystals~\cite{Bentley2013,Taylor2017,Ratcliffe2018,Mehdi2020b:2D,Wu2020b,Mehdi2021} and enables a large number of quantum logic operations to be leveraged against the long coherence times of trapped-ion qubits. However, solutions in the literature favour pulse sequences found to perform best when the ion dynamics are strongly localised~\cite{Bentley2013,Taylor2017, zouImplementationLocalHighfidelity2009}. This limits existing fast gate proposals to nearest-neighbour qubit pairs~\cite{Bentley2013,Gale2020a,Ratcliffe2018,Mehdi2021e}, which necessitates large resource overheads to realise non-local quantum logic operations using nearest-neighbour SWAPs~\cite{Taylor2017,Mehdi2020b:2D}. However, the nearest-neighbour limitation is not fundamental to the fast gate mechanism. The key result of this manuscript is the demonstration that the fast gate mechanism can be tailored to perform non-local 2Q gates between arbitrary qubit pairs in long ion chains, which paves the way for fast quantum logic in all-to-all connected ion trap modules.

\begin{figure}
\includegraphics[width=\columnwidth]{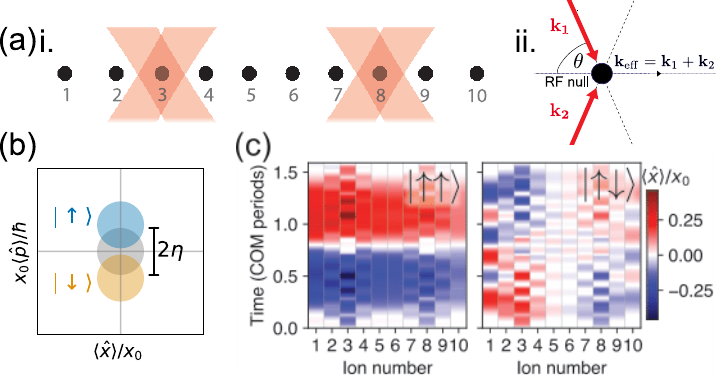}
\caption{\label{fig:Diagram} (a) Illustration of the 10-ion linear chain geometry, where axial motion of individually-addressed ions is excited using counter-propagating Raman beams tilted from the RF null by angle $\theta$ \cite{CP, Putnam2024a}. (b) The action of an SDK is a spin-dependent displacement of $\eta=k_{\rm eff}x_0$ along the momentum axis in phase space, with $k_{\rm eff}$ as the effective wavevector of the Raman pulse and $x_0=\sqrt{\hbar/(2m\omega_{\rm COM})}$. (c) Spin-dependent motion of the ion chain during a non-local fast gate (theoretical state-averaged fidelity of $99.97\%$) with $56$ SDKs addressing ions $3$ and $8$ for same-spin and different-spin 2Q states (motional trajectories for $\ket{\downarrow\downarrow}$ and $\ket{\downarrow\uparrow}$ are given by symmetry~\cite{Mehdi2025}, $\langle\hat{x}\rangle\rightarrow -\langle\hat{x}\rangle$). 
}	
\end{figure}

We identify candidate fast gate protocols between non-local qubit pairs in quasi-uniform chains of trapped-ion qubits through explicit optimisation of SDK sequences~\cite{Gale2020,CP}. We find fast non-local entangling gates between distant qubits rely on mediation by the phonon modes of the crystal, similar to conventional spin-spin interactions in trapped-ion systems~\cite{Soderberg2010}. This is illustrated in Fig.~\ref{fig:Diagram}(c), which shows the motional dynamics of a ten-ion chain during a non-local fast gate of duration $\tau_{\rm G}\approx 1.44\tau_0$, where $\tau_0=2\pi/\omega_{\rm COM}$ is a single centre-of-mass oscillation period. In this exemplary gate, all ions oscillate around their equilibrium position, which is a common feature of non-local fast gates identified in this work. This is contrasted with highly localised motion around target ions in nearest-neighbour gates in the regime $\tau_{\rm G}\ll \tau_0$~\cite{Bentley2013,Taylor2017,Mehdi2021,Wu2020b}.

\begin{figure}
\includegraphics[width=\columnwidth]{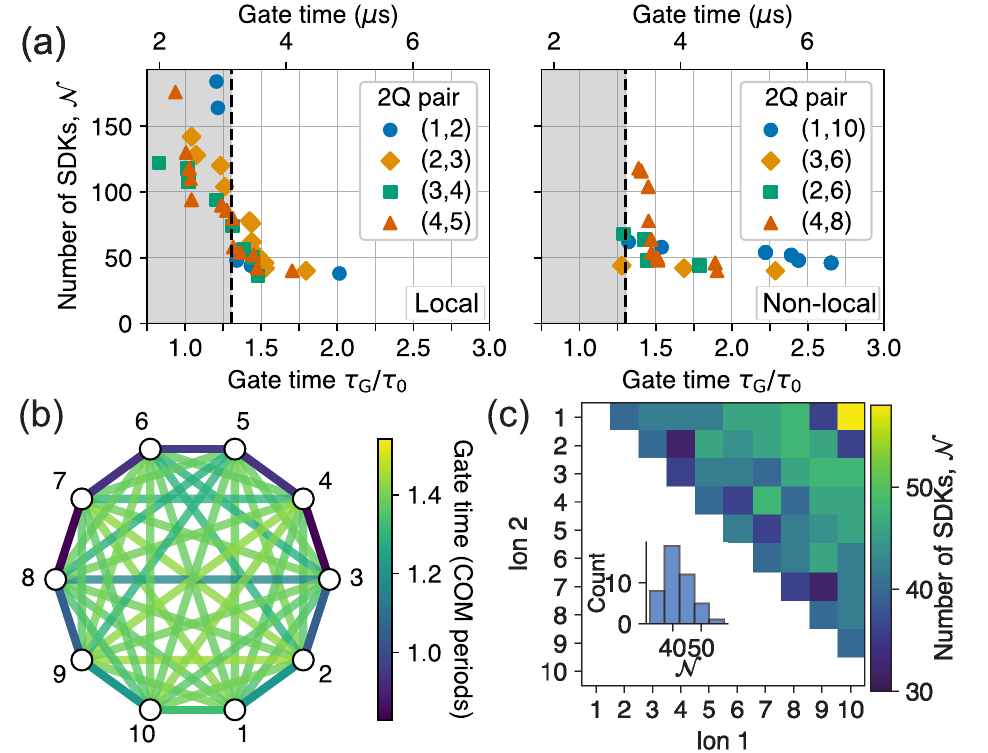}
\caption{\label{fig:10Ions_All2AllConnectivity} \textbf{Fast gates enable all-to-all connectivity in a ten-ion chain.} All gate solutions shown have a theoretical state-averaged fidelity exceeding $99.95\%$ in a ten-ion chain with inter-ion separation $\geq$  \SI{3}{\micro\meter} ($\tau_0\approx$ \SI{2.4}{\micro\second}). (a)  Number of SDKs per 2Q operation required to achieve a target gate time with theoretical fidelity above $99.95\%$ for select qubit pairings indexed by their location along the chain $(i, j)$. The vertical dotted line indicates the analytically-predicted crossover between the supersonic regime (shaded) and subsonic regime, $\tau_0/(\omega_{\rm BR}/\omega_{\rm COM}-1)\approx 1.3\tau_0$. (b) All-to-all connectivity diagram indicating the shortest gate duration achievable with fewer than $\mathcal{N}\leq 200$ SDKs per operation. (c) Minimum number of SDKs required for gates between every qubit pairing for gate operation times shorter than 2$\tau_0\approx$ \SI{5}{\micro\second}, with a histogram shown in the inset. 
}	
\end{figure}

\emph{Speed limits to phonon-mediated entanglement.---} The threshold gate time is set by the finite speed of sound in the ion chain, which sets a fundamental speed limit to phonon-mediated entangling gates in trapped-ion systems. We analytically estimate this bound by considering the time taken for a phonon wavepacket to travel the length of the chain and return to its original position. 
 That is, $\tau_{\rm G} > \tau_{\rm travel}= 2L/c_{\rm s}$, where $L$ is the length of the ion chain and $c_{\rm s}$ is the speed of sound in the chain. The latter can be estimated from the group velocity of the mode spectrum, which can be bounded as $v_g > (\omega_{\rm BR}-\omega_{\rm COM})L/\pi$ (see Appendix A), where $\omega_{\rm COM}$ and $\omega_{\rm BR}$ are the centre-of-mass and breathing (stretch) mode frequencies, respectively. In centre-of-mass units, this gives the bound $\tau_{\rm travel}\approx \tau_0/(\omega_{\rm BR}/\omega_{\rm COM}-1)$ which is in good agreement with the identified threshold in gate searches shown in Fig~\ref{fig:10Ions_All2AllConnectivity}(a) and numerical experiments tracking the position of a phonon wavepacket excited by a single impulse to an ion at the edge of the chain (see Appendix B). For a harmonic trap, $\omega_{\rm BR}=\sqrt{3}\omega_{\rm COM}$ gives $\tau_{\rm travel}\approx 1.37\tau_0$ -- the stabilising quartic term in the potential results in a slightly larger mode spacing~\cite{CP}, which reduces the value of $\tau_{\rm travel}$ to approx. $1.3\tau_0$ ($1.25\tau_0$) for a $N=10$ ($N=50$) ion chain.

Based on the above analysis, we introduce the terminology of \emph{subsonic fast gates} to describe fast gates with duration $\tau_{\rm G} > \tau_0/(\omega_{\rm BR}/\omega_{\rm COM}-1)$. Subsonic gates, such as shown in Fig.~\ref{fig:Diagram}(c), involve phonon-mediated entanglement characterised by all ions in the chain experiencing driven motion (dependent on the 2Q state of the target qubits). Conversely, we denote gates with duration $\tau_{\rm G} \ll \tau_0/(\omega_{\rm BR}/\omega_{\rm COM}-1)$ as \emph{supersonic}, to reflect the localised motion around the target ions at timescales much faster than the speed of sound in the chain. We find that supersonic fast gates are only feasible for nearest neighbour pairs, consistent with previous work~\cite{Bentley2013}. 

\emph{All-to-all connected fast gates in a ten ion chain.---} We study the feasibility of implementing fast gates on arbitrary qubit pairs in a $10$-ion chain as it is computationally feasible to perform gate searches for all ${n\choose 2}=45$ unique two-qubit pairs. 

In Appendix C, we describe the heuristic approach used to find high-fidelity fast gate solutions. In our solutions we assume a quasi-uniform chain of $^{133}$Ba ions in an anharmonic axial trapping potential, $V_{\rm{trap}}(z) = \kappa_{2}z^{2}/2 + \kappa_{4}z^{4}/4$~\cite{Lin2009, CP}, that enforces an inter-ion spacing $d \geq$~\SI{3}{\micro\meter}~\cite{CP, Chen2024, pogorelovCompactIonTrapQuantum2021}. We consider impulsive state-dependent kicks (SDKs) implemented using pairs of \SI{532}{\nano\meter} Raman beams coupled to the axial modes of the crystal with an effective wave-vector $k_{\rm{eff}} = \sqrt{2}k$ ~\cite{Putnam2024a, CP} (see Fig.~\ref{fig:Diagram}(a)). In all solutions, the timing between consecutive SDKs is constrained to $|t_{k} - t_{k+1}| \geq$ \SI{10}{\nano\second} for compatibility with Raman SDK implementations~\cite{Campbell2010a,Mizrahi2013b,Johnson2015,Wong-Campos2017a, liuHighFidelityRamanSpinDependent2025}. We assume the axial modes are cooled to $T=$ \SI{30}{\micro\kelvin}, which can be realised using sub-Doppler cooling techniques~\cite{fengEfficientGroundStateCooling2020, wuElectromagneticallyinducedtransparencyCoolingTripod2025} and has contributions to the residual spin-motional entanglement error~\cite{CP}. 

The analysis in this section is restricted to candidate fast gate solutions with infidelity below $5\times 10^{-4}$ (before inclusion of pulse errors), which includes error contributions from residual spin-motional entanglement, and imperfect two-qubit phase accumulation~\cite{CP,Mehdi2025}. We characterise the best solutions by those that minimise the total gate duration $\tau_{\rm G}$ as well as the total number of SDKs $\mathcal{N}$. While these solutions are not expected to be the global optimum due to undersampling of the high-dimensional search space, we find the observed trends are robust to increases in the sampling density. The family of optimal solutions is characterised by a Pareto front in the parameter space $(\tau_{\rm G},\mathcal{N})$ (Fig.~\ref{fig:10Ions_All2AllConnectivity}(a)), which physically represents the trade off between gate speed and available laser resources.

Figure~\ref{fig:10Ions_All2AllConnectivity} illustrates the key result of this section: any two-qubit pair in the ten ion chain can be entangled with high-fidelity using (subsonic) fast gates with duration $1.4\lesssim \tau_{\rm G}/\tau_0\lesssim 2$ using tens of SDKs. For nearest-neighbour pairings, Fig.~\ref{fig:10Ions_All2AllConnectivity}(a) shows high-fidelity gate solutions exist in both the supersonic ($\tau_{\rm G}<\tau_{\rm travel}$) and subsonic ($\tau_{\rm G}>\tau_{\rm travel}$) regimes, with the fewest pulses required in the subsonic regime. For arbitrary non-local qubit pairings, we find solutions exist in the subsonic regime with $\mathcal{N}\lesssim 50$ SDKs. This is highlighted in the all-to-all connectivity diagram in Fig.~\ref{fig:10Ions_All2AllConnectivity}(b). Fig.~\ref{fig:10Ions_All2AllConnectivity}(c) shows that, for any two-qubit pair in the ten ion system, high-fidelity entangling gates can be implemented in less than two centre-of-mass periods with $30-60$ SDKs.

\begin{figure}
	\includegraphics[width=\columnwidth]{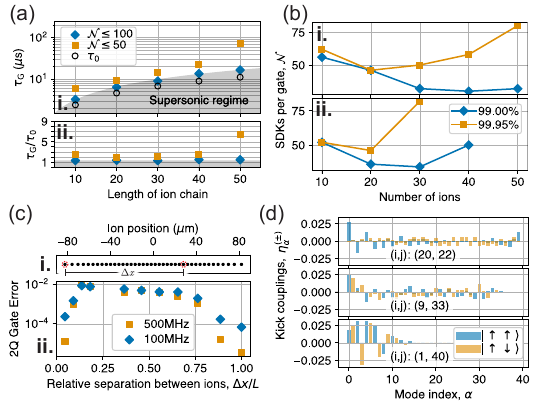}
	\caption{ 
	\label{fig:ScalingResults}
	\textbf{Subsonic fast gate scaling with length of ion chain.}  (a) 2Q gate operation time between qubits at either end of the ion chain, with state-averaged theoretical fidelity above $99.95\%$ ($\varepsilon_{\rm av}\leq 5\times 10^{-4}$) and limited number of SDKs per gate ($\mathcal{N}$). (a.i) Gate time in microseconds for an ion chain with minimum equilibrium separation of $d=$ \SI{3}{\micro\meter}, with open black circles indicating the centre-of-mass period $\tau_0=2\pi/\omega_{\rm COM}$. Gate duration in units of centre-of-mass periods is shown in (a.ii). 
	(b) Number of SDKs required to reach a target theoretical fidelity as a function of chain length for qubit pairs i. at either end of the ion chain; and ii.  separated by two intermediary ions around the trap centre. (c) Theoretical 2Q gate error for a $40$ ion chain using subsonic fast gates with operation times shorter than $2\tau_0$ ($\approx$ \SI{18}{\micro\second} for $d=$ \SI{3}{\micro\meter}), as a function of separation between the target qubits. Results are given for ion pairs symmetric about the trap centre, using an effective SDK repetition rate of \SI{100}{\mega\hertz} (blue diamonds) and \SI{500}{\mega\hertz} (orange squares). (d) Coupling of the SDKs on various qubit pairs, indexed by $(i,j)$, to the normal modes of a 40-ion chain. }
\end{figure}

\emph{Scaling to longer chains.---} Next, we investigate the scalability of the fast gate mechanism as the number of ions in the chain is increased while retaining an inter-ion separation of $d\gtrapprox$ \SI{3}{\micro\meter} ~\cite{CP}). We are interested in three key metrics: gate duration ($\tau_{\rm G}$), number of SDKs per gate ($\mathcal{N}$), and achievable fidelity of the fast gate. 

Figure \ref{fig:ScalingResults}(a) shows the achievable 2Q gate duration between qubits at either end of the ion chain with theoretical fidelity above $99.95\%$, for different chain lengths up to $N=50$ ions. We show that, for this non-local 2Q pair, gate times close to the fundamental limit for subsonic gates -- $\tau_{\rm G}\approx 1.3\tau_0 \sim N^{0.9}$~\cite{Cetina2022} -- can be achieved using tens of SDKs, each separated by at least \SI{10}{\nano\second}. For the subset of gate solutions that meet stronger constraints on the number of SDKs, slightly longer gate times ($2-10\tau_0$) are required to find high-fidelity solutions: e.g. for $\mathcal{N}\leq 50$, Fig~\ref{fig:ScalingResults}(a) shows gate durations of approx. $3\tau_0$ ($7\tau_0$) are required with $N=30$ ($N=50$) ions.

We further investigate the dependence of $\mathcal{N}$ (i.e. laser control requirements) with chain length in Fig.~\ref{fig:ScalingResults}(b), where we compare subsonic gates with duration $\leq 2\tau_0$ between qubit pairs at i.~either end of the chain; and ii.~separated by a fixed distance of $\approx$ \SI{30}{\micro\meter} (eight ions apart) about the centre of the chain. 

The required number of SDKs decreases with chain length for gates with theoretical fidelity above $99\%$ up to $N=50$ ion chains. 
For end-to-end gates (case 1), we find that the number of SDKs required to achieve gate fidelities above $99.9\%$ modestly increases from $\mathcal{N}\approx50$ in a twenty ion chain to $\mathcal{N}\approx 80$ in a fifty ion chain. We observe similar behaviour for the ion pair with fixed separation (case 2), however we are unable to find gate solutions with theoretical fidelity above $99\%$ ($99.9\%$) for chains longer than $N=40$ ($N=30$) ions. This indicates a limit to the scalability of subsonic (nonlocal) fast gates in long ion chains, which further depends on the specific ion pair being addressed.

In Figure \ref{fig:ScalingResults}(c) we investigate the dependence on achievable gate fidelity between ion pairs in an $N=40$ ion chain with varying inter-ion separation between approx. \SI{3}{\micro\meter} (nearest neighbours) and $L=$ \SI{164}{\micro\meter} (ions at either end of the chain), where $L$ is the length of the chain. We restrict our consideration to subsonic gates with duration $\leq 2\tau_0\approx$ \SI{18}{\micro\second}. Fig.~\ref{fig:ScalingResults}(c.ii) shows the theoretical gate error is as large as $1\%$ for inter-ion separations $\Delta x$ in the approximate range $10\%<\Delta x/L<75\%$. The smallest gate errors achievable are at the $10^{-4}$ level for nearest-neighbour pairs (consistent with previous work~\cite{Mehdi2021e}) and for pairs with the ions near opposite ends of the chain. Increasing the timing freedoms of the pulse sequences by reducing the minimum separation between SDKs from \SI{10}{\nano\second} to \SI{2}{\nano\second} (i.e. increasing the effective SDK repetition rate from \SI{100}{\mega\hertz} to \SI{500}{\mega\hertz}) improves the gate fidelity between distant ions, however does not improve the gate fidelity for most ion pairings.

The scaling of nonlocal fast gates in long ion chains is limited by residual spin-motional entanglement errors, which increase in larger crystals with more motional modes. This can be understood by considering the coupling of SDKs on an ion pair $(i,j)$ to the $\alpha$-th motional mode which is proportional to $\eta_\alpha\left(b_\alpha^{(i)}\hat{\sigma}_z^{(i)}+b_\alpha^{(j)}\hat{\sigma}_z^{(j)}\right)$ where $b_\alpha^{(i)}$ is the mode eigenvector~\cite{CP,Mehdi2025}, shown in Fig.~\ref{fig:ScalingResults}(d) for select pairings in a $40$ ion chain. For select pairings such as the boundary ions $(i=1,j=N)$, only a subset of low-frequency motional modes are excited by the kicks. This effectively reduces the number of motional restoration conditions that must be satisfied by the SDK sequence, which enables higher achievable fidelities with a limited number of SDKs. 

With the available freedoms in our gate search (the timing and direction of each SDK, for $\mathcal{N}\lesssim 10^{2}$), we find high-fidelity subsonic gates between arbitrary pairs become infeasible for chains longer than $40$ ions. In principle these schemes could be extended to longer crystals by tuning mode couplings via motional mode engineering~\cite{Li2025,Cheng2024}, or by adding more freedoms to the pulse sequence at the cost of requiring a larger number of SDKs and improved pulse control~\cite{Gale2020a}. Given coherent control of individually-addressed ion qubits is challenging beyond $50$ ions due to thermal motion of the ions within focused laser beams~\cite{Cetina2022}, we do not explore these possibilities in this manuscript.

\emph{Experimental feasibility.---} The fast gate mechanism can be applied to a wide range of ion species and qubit types, including mixed-species crystals~\cite{Mehdi2025}, constrained only by the anharmonicity of the trap, and the ability to perform high-fidelity SDKs on individually addressed ions.

Coupling to the axial modes of individually addressed ions poses a challenge for long ion chains. For the two-photon Raman SDKs considered in this work, individual ion addressing is possible using a pair of tightly-focused beams equally tilted from the RF null of the trap as shown in Fig.~\ref{fig:Diagram}(a). Alternatively, in single-photon SDKs, this constraint can also be satisfied using structured light~\cite{cuiTransversePolarizationGradient2025, maiScalableEntanglingGates2025} or by shelving all non-target ions in off-resonant states~\cite{Landsman2019}.
We assume the thermal motion of the ions is sufficiently small to enable individual addressing, which is reasonable for the $T=$~\SI{30}{\micro\kelvin} assumed in this work. We also assume the ion separation is large enough to prevent unwanted excitations of non-targeted ions, which is achievable for the $d \geq$ \SI{3}{\micro\meter} ion spacing enforced in this work~\cite{Chen2024, pogorelovCompactIonTrapQuantum2021}. High-fidelity solutions can also be found for larger ion separations, at the cost of longer gate times given the centre-of-mass period scales with the ion separation as $\tau_{0} \propto d^{\frac{3}{2}}$. For example, $\tau_{0}=$~\SI{5}{\micro\second} in a ten-ion linear crystal with $d \geq$~\SI{5}{\micro\meter}.

The primary technical limitation to fast gate implementation is achieving high fidelity SDKs. There has only been a single experimental fast gate demonstration~\cite{Wong-Campos2017a} with a reported fidelity of $76\%$ limited by compounding pulse errors~\cite{Bentley2016a,Gale2020a}. In the companion manuscript~\cite{CP}, we analyse the contributions of population inversion errors and motional diffraction errors from imperfect SDKs to the total gate error. Providing the former can be suppressed (e.g. by engineering a $\pi$ phase shift between consecutive SDKs), we find the schemes in this work can be implemented with $95\%$ fidelities with current SDK fidelities of $\sim 99\%$~\cite{johnsonUltrafastCreationLarge2017, hussainUltrafastHighRepetition2016,Johnson2015}. Achieving $99.95\%$ 2Q fidelities requires further suppression of SDK errors to the $10^{-4}$ level which is comparable to single-qubit fidelities demonstrated on trapped-ion hardware, however is yet to be achieved at nanosecond timescales. However, the broad range of demonstrated SDK implementation schemes~\cite{Mizrahi2013a,Johnson2015, hussainUltrafastHighRepetition2016, Heinrich2019a, Guo2022b, Putnam2024a, Campbell2010a} provides a pathway to overcoming pulse error limitations in near-future experiments. Furthermore, recent proposals suggest Raman-SDK errors can be suppressed to the spontaneous emission limit at $10^{-7}$ using modulated continuous-wave lasers~\cite{liuHighFidelityRamanSpinDependent2025}. Alternative SDK implementations such as using fast-switched electrodes and a global SDK beam in combination with Rydberg excitations ~\cite{Zhang2020b, baoQuantumComputingArchitecture2025} may also provide further improvements in future work.

\emph{Conclusion.---} In this Letter, we have expanded existing fast 2Q gate protocols to enable high-fidelity gates between non-local qubit pairs in linear ion chains, indicating all-to-all connected quantum logic rates of $\mathcal{O}(100)$kHz are achievable in systems with tens of qubits. This represents a potential hundred-fold increase in quantum processing rates over current demonstrations with linear chains of up to $30$ ions~\cite{Chen2024}, and provides a clear path toward executing high-depth quantum circuits with $\mathcal{O}(10^6)$ two-qubit operations within the long coherence times ($>$ \SI{10}{\second}) of hyperfine trapped-ion qubits. Using established fast gate optimisation algorithms we identified high-fidelity gate solutions between arbitrary ion pairs in chains up to $40$ ions. Notably, the pulse requirements (i.e. number of SDKs per 2Q operation) do not increase with chain length in contrast to conventional gate mechanisms such as the M\o lmer–S\o rensen protocol~\cite{Sorensen2000}, where the coupling to the target motional mode scales as $1/\sqrt{N}$.

Given the control challenges present in larger ion chains, scaling beyond $\sim 40$ qubits will require integrating the subsonic fast gate mechanism into modular-based architectures. A natural pathway for implementing this is to use chains of $40$ ions which can then be connected using photonic interfaces~\cite{Monroe2014,Drmota2023a,Main2025,OReilly2024}, shuttling and recombination of ion chains ~\cite{Moses2023,Mordini2025,Akhtar2023}, or inter-trap fast gates~\cite{Ratcliffe2018,Mehdi2020b:2D}.  This is advantageous as inter-trap connections via shuttling or photonic interconnects are orders of magnitude slower than (intra-trap) fast 2Q gates~\cite{Mehdi2020b:2D}, allowing deeper circuit implementations without bottlenecking the absolute computation speed. Future work could also extend fast gate design to multi-dimensional ion crystals.

The data supporting the findings of this Letter is openly available~\cite{mehdiDatasetHighSpeed2025a}.

\emph{Acknowledgments.---} This research was undertaken with the assistance of supercomputing resources and services from the National Computational Infrastructure, which is supported by the Australian Government. ISB thanks the support provided by the D.N.F Dunbar Honours Scholarship and the Australian Government Research Training Program (RTP) Scholarship.

\pagebreak
\appendix

\begin{center}
    \textbf{Appendices}
\end{center}

\section{A. Derivation of round-trip travel time of a phononic wavepacket}
Here we provide a derivation of the phononic sound cone in a uniformly spaced $N$-ion chain, with equilibrium positions $x_0^{(j)} = j d$ along the RF null of the trap. 

The motional mode eigenvectors are given by plane waves with (anti-)nodes at the equilibrium positions of the ions, i.e. $b_n^j =\exp\left(ik_n x^{(j)}_0\right)/\sqrt{N}$ where $k_n = n\pi/L$ with $L$ as the length of the chain.

The group velocity of a phonon wavepacket is given by $v_g = \partial\omega(k)/\partial k$. In order to apply this to an ion chain of finite length with a discrete mode spectrum, we apply the chain rule and use a finite difference between mode frequencies, i.e. $\partial \omega/\partial k \approx (\Delta \omega/\Delta n) \partial n/\partial_k$. Using $\partial k/\partial n = \pi/L$ and $\Delta n = 1$, we then have the expression for the group velocity
\begin{align}
	v_{g,n} \approx\frac{\Delta \omega_n L}{\pi}\,,
\end{align}
where $\Delta \omega_n = \omega_{n+1}-\omega_n$ is the $n$-th mode splitting. As the mode splittings are not equal, the group velocity is not constant and depends on the spectrum of the system.

We are primarily interested in deriving a bound to the speed at which phonon wavepackets can propagate along the chain. It is therefore sufficient to bound the group velocity by the largest mode splitting, which is between the lowest two frequency modes -- the centre-of-mass oscillation ($\omega_{\rm COM}$) and the breathing/stretch mode ($\omega_{\rm BR}$) -- i.e.
\begin{align}
\label{eq:GroupVelocity_bound}
	v_g \lessapprox \frac{L(\omega_{\rm BR}-\omega_{\rm COM})}{\pi}\,.
\end{align}

We can thus bound the time taken for a phonon wavepacket to traverse twice the length of the ion chain as:
\begin{align}
\label{eq:TravelTime_Analytic}
	\tau_{\rm travel} \gtrapprox \frac{2 L}{{\rm max}(v_g)}=\frac{1}{\omega_{\rm BR}/\omega_{\rm COM}-1} \tau_0\,.
\end{align}
For a harmonic trap, $\omega_{\rm BR} = \sqrt{3}\omega_{\rm COM}$, which gives $\tau_{\rm travel} \lessapprox 1.37\tau_0$. The addition of a quartic term in the trapping potential, which is necessary to ensure quasi-uniform spacing~\cite{Home2011}, leads to a slightly larger value of $\omega_{\rm BR}/\omega_{\rm COM}$. This shift depends on the number of ions in the chain -- for $N=30$ ions, we have $\omega_{\rm BR}/\omega_{\rm COM}\approx 1.8$ which gives $\tau_{\rm travel}\lessapprox 1.26\tau_0$.

\subsection{B. Numerical experiment: Phononic sound cone in a $30$ ion chain}
\begin{figure}
\includegraphics[width=\columnwidth]{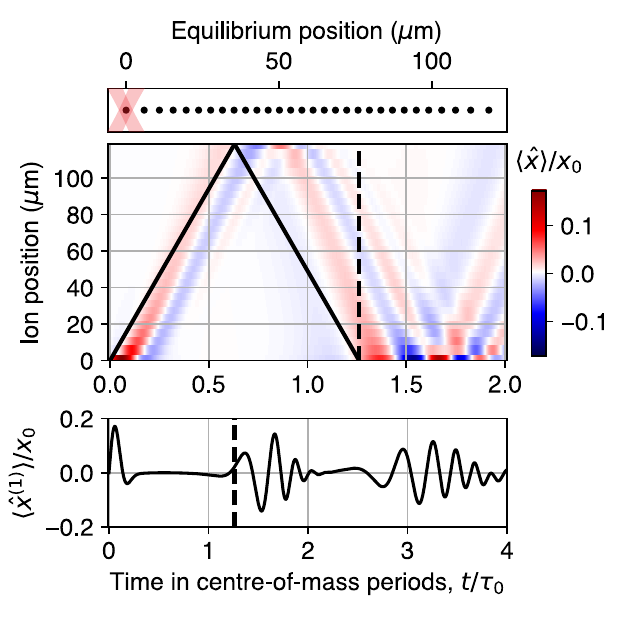}
	\caption{ 
	\label{fig:NumericalExperiment}
	\textbf{Numerical experiment demonstrating speed limits to phonon-mediated interactions between distant ion qubits.} The solid black line indicates the phononic sound cone with slope given by the analytically-estimated group velocity, Eq.~\eqref{eq:GroupVelocity_bound}, and the dashed black line indicates the estimated round-trip travel time Eq.~\eqref{eq:TravelTime_Analytic}. The bottom figure shows the dimensionless displacement of the initially-kicked ion as a function of time; the beginning of the revival due to the return of the wavepacket to ion $1$ is in excellent agreement with the analytic prediction, Eq.~\eqref{eq:TravelTime_Analytic}.    }
\end{figure}
Here we demonstrate the validity of the analytically estimated minimum travel time for a phononic wavepacket, Eq.~\eqref{eq:TravelTime_Analytic}, by a direct numerical experiment in a $30$-ion chain. In the simulated scenario, an ion at the boundary of the chain (ion $1$) is subjected to an impulsive momentum kick at $t=0$, and the resulting displacement of the ion chain is tracked over several centre-of-mass oscillation periods. The results of this calculation are shown in Figure~\ref{fig:NumericalExperiment}, which illustrates the impulsive perturbation on the boundary ion as it propagates through the ion chain, with its position accurately estimated by the group velocity Eq.~\eqref{eq:GroupVelocity_bound}. We find that Eq.~\eqref{eq:TravelTime_Analytic} correctly predicts the time at which the perturbation begins to return to the initially kicked ion, as illustrated by a revival in the displacement of ion $1$ from its equilibrium position.

\section{C. Gate design}

In this Appendix, we summarise the details of the numerical optimisation approach we adopt to design gate schemes. Further details on these optimisation methods can be found in the companion manuscript~\cite{CP}. 

To realise a maximally entangling $\hat{\sigma}_{z} \bigotimes \hat{\sigma}_{z}$ phase gate between a pair of ions in an $N$-ion chain, the SDK sequence must satisfy a set of $N+1$ conditions which correspond to the restoration of the ion motion at the end of the gate, and the accumulation of the correct state-dependent phase. This is a high-dimensional problem with no analytical solution so we take a heuristic approach to find the sequence of SDKs that implement a high-fidelity fast gate.

For optimisation of the number and timing of SDKs we use an anti-symmetric gate (APG) scheme composed of $K$ groups of SDKs~\cite{Gale2020a}:
\begin{align}
	\bm{z} &= \{-z_{K/2},...-z_{2},-z_{1}, z_{1}, z_{2}, ..., z_{K/2}\}~,\\
	\bm{t} &= \{-t_{K/2}, ..., -t_{2}, -t_{1}, t_{1},t_{2},..., t_{K/2}\} ~,
\end{align}
where the $j$-th group is composed of $z_{j}$ SDKs centred around $t_{j}$ and the sign of $z_{j}$ corresponds to the direction of the SDKs in that group.

We use a truncated expression for the state-averaged infidelity of a gate between ions $A$ and $B$ in terms of the 2Q phase-acquisition error ($\Delta \Theta$) and residual spin-motional entanglement ($\Delta \beta$) as our cost function~\cite{CP}:
\begin{align}
    \varepsilon_{\rm{av}} = \frac{2}{3}& \left\lvert\Delta \Theta\right\rvert^{2}\notag\\& + \frac{4}{3} \sum_{\alpha=1}^{N} \left(\frac{1}{2} + \bar{n}_{\alpha}\right)\left[(b_{\alpha}^{(A)})^{2} + (b_{\alpha}^{(B)})^{2}\right] \left\lvert\Delta \beta_{\alpha}\right\rvert^{2}~,\label{eq:infidelity}
\end{align}
where $b_{\alpha}^{(A)}$ is the coupling of the $\alpha$-th normal mode to the $A$-th ion and $\bar{n}_{\alpha} = \left( e^{\frac{\hbar \omega_{\alpha}}{k_{B}T}} -1\right)^{-1}$ is the average phonon occupation of $\alpha$-th mode with oscillation frequency $\omega_{\alpha}$ in terms of the temperature of the system, $T$. We assume the system temperature is constant throughout the gate given the gates we consider are much faster than reported trap heating rates~\cite{bruzewiczMeasurementIonMotional2015, paganoCryogenicTrappedIonSystem2018, Harty2014}.

In this expression, the error in the 2Q phase-acquisition is given by
\begin{equation}\label{eq:entangling_phase_condition}
    \Delta\Theta = \left\lvert2\sum_{\alpha = 1}^{N}\eta_{\alpha}^{2} b_{\alpha}^{(m)}b_{\alpha}^{(n)}\sum_{k\neq j} ^{\mathcal{N}}z_{j}z_{k}\sin\left(\omega_{\alpha}(t_{j} - t_{k})\right)\right\rvert - \frac{\pi}{4}
\end{equation}
and the residual spin-motional entanglement for the $\alpha$-th mode is, 
\begin{equation}
     \Delta \beta_{\alpha} = i\eta_{\alpha}\sum_{j=1}^{\mathcal{N}} z_{j}\sin(\omega_{\alpha} t_{j})~.\label{eq:motional_displacement_condition_APG}
\end{equation}
We note that this expression is a simplified version of the more general final spin-motional entanglement $\Delta \beta_{\alpha} = i\eta_{\alpha}\sum_{j=1}^{\mathcal{N}} z_{j} e^{i\omega_{\alpha} t_{j}}$ as the APG scheme guarantees momentum restoration for each mode.

The mode couplings and frequencies for each chain length are calculated by performing a normal mode expansion of the ion displacements from equilibrium assuming the effective potential experienced by each ion can be truncated at second order~\cite{James1998a}. Consequently, anharmonicities in the total potential will impact the gate fidelities we report, contributing errors on the order of $10^{-6}$ to the total gate error~\cite{Gale2020}.

Each optimisation is performed in two stages where the gate error is minimised under constraints on the number of SDKs, and the gate time. In the first stage, the number of SDK groups and the maximum number of SDKs per group is bounded and a global optimisation is performed over $\mathbf{z}$ while the timings are fixed based on a desired gate time $\tau_{G}$: $\mathbf{t} = \frac{\tau_{G}}{K}\{-\frac{K}{2}, ..., -2, -1, 1,2,..., \frac{K}{2}\}$. The optimal solutions are then used in the second optimisation stage, where local optimisations are performed on $\bm{t}$ to further refine the gate solution and  a minimum timing separation is enforced to separate SDKs within each group, $t_{j+1} = t_{j} + 1/f_{\rm{rep}}$, for a specific SDK repetition rate, $f_{\rm{rep}}$.
 
\bibliographystyle{bibsty}
\bibliography{bib}

@article{hussainUltrafastHighRepetition2016,
  title = {Ultrafast, High Repetition Rate, Ultraviolet, Fiber-Laser-Based Source: Application towards {{Yb}}{\textsuperscript{+}} Fast Quantum-Logic},
  shorttitle = {Ultrafast, High Repetition Rate, Ultraviolet, Fiber-Laser-Based Source},
  author = {Hussain, Mahmood Irtiza and Petrasiunas, Matthew Joseph and Bentley, Christopher D. B. and Taylor, Richard L. and Carvalho, Andr{\'e} R. R. and Hope, Joseph J. and Streed, Erik W. and Lobino, Mirko and Kielpinski, David},
  year = {2016},
  month = jul,
  journal = {Optics Express},
  volume = {24},
  number = {15},
  pages = {16638--16648},
  publisher = {Optica Publishing Group},
  issn = {1094-4087},
  doi = {10.1364/OE.24.016638},
  urldate = {2024-08-25},
  copyright = {{\copyright} 2016 Optical Society of America},
  langid = {english}
}

@article{cuiTransversePolarizationGradient2025,
  title = {Transverse {{Polarization Gradient Entangling Gates}} for {{Trapped-Ion Quantum Computation}}},
  author = {Cui, Jin-Ming and Chen, Yan and Zhou, Yi-Fan and Long, Quan and An, En-Teng and He, Ran and Huang, Yun-Feng and Li, Chuan-Feng and Guo, Guang-Can},
  year = 2025,
  month = dec,
  journal = {Physical Review Letters},
  volume = {135},
  number = {26},
  eprint = {2506.19691},
  primaryclass = {quant-ph},
  pages = {260604},
  issn = {0031-9007, 1079-7114},
  doi = {10.1103/w5l6-wmrl},
  urldate = {2026-01-23},
  archiveprefix = {arXiv}
}

@article{pogorelovCompactIonTrapQuantum2021,
  title = {Compact {{Ion-Trap Quantum Computing Demonstrator}}},
  author = {Pogorelov, I. and Feldker, T. and Marciniak, {\relax Ch}. D. and Postler, L. and Jacob, G. and Krieglsteiner, O. and Podlesnic, V. and Meth, M. and Negnevitsky, V. and Stadler, M. and H{\"o}fer, B. and W{\"a}chter, C. and Lakhmanskiy, K. and Blatt, R. and Schindler, P. and Monz, T.},
  year = 2021,
  month = jun,
  journal = {PRX Quantum},
  volume = {2},
  number = {2},
  pages = {020343},
  issn = {2691-3399},
  doi = {10.1103/PRXQuantum.2.020343},
  urldate = {2025-11-17},
  langid = {english}
}

@misc{maiScalableEntanglingGates2025,
  title = {Scalable Entangling Gates on Ion Qubits via Structured Light Addressing},
  author = {Mai, Xueying and Zhang, Liyun and Yu, Qinyang and Zhang, Junhua and Lu, Yao},
  year = 2025,
  month = jun,
  number = {arXiv:2506.19535},
  eprint = {2506.19535},
  primaryclass = {quant-ph},
  publisher = {arXiv},
  doi = {10.48550/arXiv.2506.19535},
  urldate = {2025-06-28},
  archiveprefix = {arXiv}
}

@article{baoQuantumComputingArchitecture2025,
  title = {Quantum Computing Architecture with Trapped Ion Crystals and Fast {{Rydberg}} Gates},
  author = {Bao, Han and Vogel, Jonas and Poschinger, Ulrich and {Schmidt-Kaler}, Ferdinand},
  year = 2025,
  month = apr,
  journal = {Physical Review Research},
  volume = {7},
  number = {2},
  pages = {023035},
  publisher = {American Physical Society},
  doi = {10.1103/PhysRevResearch.7.023035},
  urldate = {2025-07-14}
}

@misc{liuHighFidelityRamanSpinDependent2025,
  title = {High-{{Fidelity Raman Spin-Dependent Kicks}} in the {{Presence}} of {{Micromotion}}},
  author = {Liu, Haonan and Vaidya, Varun D. and Galan, Monica Gutierrez and Ratcliffe, Alexander K. and Poudel, Amrit and Viteri, C. Ricardo},
  year = 2025,
  month = nov,
  number = {arXiv:2511.15959},
  eprint = {2511.15959},
  primaryclass = {quant-ph},
  publisher = {arXiv},
  doi = {10.48550/arXiv.2511.15959},
  urldate = {2025-12-06},
  archiveprefix = {arXiv}
}

@misc{zouImplementationLocalHighfidelity2009,
  title = {Implementation of Local and High-Fidelity Quantum Conditional Phase Gates in a Scalable Two-Dimensional Ion Trap},
  author = {Zou, Ping and Xu, Jian and {song}, Wei and Zhu, Shi-Liang},
  year = {2009},
  month = nov,
  eprint = {0906.4598},
  primaryclass = {quant-ph},
  doi = {10.1016/j.physleta.2010.01.035},
  urldate = {2025-07-15},
  archiveprefix = {arXiv}
}

@article{wuElectromagneticallyinducedtransparencyCoolingTripod2025,
  title = {Electromagnetically-Induced-Transparency Cooling with a Tripod Structure in a Hyperfine Trapped Ion with Mixed-Species Crystals},
  author = {Wu, Jenny J. and Hou, Pan-Yu and Erickson, Stephen D. and Brandt, Adam D. and Wan, Yong and Zarantonello, Giorgio and Cole, Daniel C. and Wilson, Andrew C. and Slichter, Daniel H. and Leibfried, Dietrich},
  year = {2025},
  month = apr,
  journal = {Physical Review A},
  volume = {111},
  number = {4},
  pages = {043109},
  issn = {2469-9926, 2469-9934},
  doi = {10.1103/PhysRevA.111.043109},
  urldate = {2025-08-06},
  langid = {english}
}

@article{fengEfficientGroundStateCooling2020,
  title = {Efficient {{Ground-State Cooling}} of {{Large Trapped-Ion Chains}} with an {{Electromagnetically-Induced-Transparency Tripod Scheme}}},
  author = {Feng, L. and Tan, W. L. and De, A. and Menon, A. and Chu, A. and Pagano, G. and Monroe, C.},
  year = {2020},
  month = jul,
  journal = {Physical Review Letters},
  volume = {125},
  number = {5},
  pages = {053001},
  issn = {0031-9007, 1079-7114},
  doi = {10.1103/PhysRevLett.125.053001},
  urldate = {2025-08-06},
  langid = {english}
}

@article{paganoCryogenicTrappedIonSystem2018,
  title = {Cryogenic {{Trapped-Ion System}} for {{Large Scale Quantum Simulation}}},
  author = {Pagano, G. and Hess, P. W. and Kaplan, H. B. and Tan, W. L. and Richerme, P. and Becker, P. and Kyprianidis, A. and Zhang, J. and Birckelbaw, E. and Hernandez, M. R. and Wu, Y. and Monroe, C.},
  year = {2018},
  month = oct,
  journal = {Quantum Science and Technology},
  volume = {4},
  number = {1},
  eprint = {1802.03118},
  primaryclass = {physics, physics:quant-ph},
  pages = {014004},
  issn = {2058-9565},
  doi = {10.1088/2058-9565/aae0fe},
  urldate = {2024-09-03},
  archiveprefix = {arXiv},
  langid = {english}
}

@article{zhuTrappedIonQuantum2006,
  title = {Trapped {{Ion Quantum Computation}} with {{Transverse Phonon Modes}}},
  author = {Zhu, Shi-Liang and Monroe, C. and Duan, L.-M.},
  year = {2006},
  month = aug,
  journal = {Physical Review Letters},
  volume = {97},
  number = {5},
  pages = {050505},
  issn = {0031-9007, 1079-7114},
  doi = {10.1103/PhysRevLett.97.050505},
  urldate = {2024-08-22},
  copyright = {http://link.aps.org/licenses/aps-default-license},
  langid = {english}
}

@article{johnsonUltrafastCreationLarge2017,
  title = {Ultrafast Creation of Large {{Schr{\"o}dinger}} Cat States of an Atom},
  author = {Johnson, K. G. and {Wong-Campos}, J. D. and Neyenhuis, B. and Mizrahi, J. and Monroe, C.},
  year = {2017},
  month = sep,
  journal = {Nature Communications},
  volume = {8},
  number = {1},
  pages = {697},
  publisher = {Nature Publishing Group},
  issn = {2041-1723},
  doi = {10.1038/s41467-017-00682-6},
  urldate = {2025-06-23},
  copyright = {2017 The Author(s)},
  langid = {english}
}

@dataset{mehdiDatasetHighSpeed2025a,
  title = {Dataset for "{{High}} Speed and High Connectivity Two-Qubit Gates in Trapped-Ion Processors"},
  author = {Mehdi, Zain and Savill-Brown, Isabelle},
  date = {2025-08-13},
  publisher = {Zenodo},
  url = {https://zenodo.org/records/16824480},
  urldate = {2025-08-13}
}

@article{bruzewiczMeasurementIonMotional2015,
  title = {Measurement of Ion Motional Heating Rates over a Range of Trap Frequencies and Temperatures},
  author = {Bruzewicz, C. D. and Sage, J. M. and Chiaverini, J.},
  year = {2015},
  month = apr,
  journal = {Physical Review A},
  volume = {91},
  number = {4},
  pages = {041402},
  issn = {1050-2947, 1094-1622},
  doi = {10.1103/PhysRevA.91.041402},
  urldate = {2024-10-11},
  copyright = {http://link.aps.org/licenses/aps-default-license},
  langid = {english}
}

@article{CP,
  title = {Error-Resilient Fast Entangling Gates for Scalable Ion-Trap Quantum Processors},
  author = {{Savill-Brown}, Isabelle and Mehdi, Zain and Ratcliffe, Alexander K. and Vaidya, Varun D. and Liu, Haonan and Haine, Simon A. and Viteri, C. Ricardo and Hope, Joseph J.},
  year = 2026,
  month = may,
  journal = {Physical Review A},
  volume = {113},
  number = {5},
  pages = {052610},
  publisher = {American Physical Society},
  doi = {10.1103/4thn-7wyf},
  urldate = {2026-05-15}
}

@article{Lin2009,
  title = {Large-scale quantum computation in an anharmonic linear ion trap},
  volume = {86},
  ISSN = {1286-4854},
  url = {http://dx.doi.org/10.1209/0295-5075/86/60004},
  DOI = {10.1209/0295-5075/86/60004},
  number = {6},
  journal = {EPL (Europhysics Letters)},
  publisher = {IOP Publishing},
  author = {Lin,  G.-D. and Zhu,  S.-L. and Islam,  R. and Kim,  K. and Chang,  M.-S. and Korenblit,  S. and Monroe,  C. and Duan,  L.-M.},
  year = {2009},
  month = jun,
  pages = {60004}
}

@article{Akhtar2023,
  title = {A High-Fidelity Quantum Matter-Link between Ion-Trap Microchip Modules},
  author = {Akhtar, M. and Bonus, F. and {Lebrun-Gallagher}, F. R. and Johnson, N. I. and {Siegele-Brown}, M. and Hong, S. and Hile, S. J. and Kulmiya, S. A. and Weidt, S. and Hensinger, W. K.},
  year = {2023},
  month = feb,
  journal = {Nat Commun},
  volume = {14},
  number = {1},
  pages = {531},
  publisher = {Nature Publishing Group},
  issn = {2041-1723},
  doi = {10.1038/s41467-022-35285-3},
  urldate = {2025-05-15},
  copyright = {2023 The Author(s)},
  langid = {english}
}

@article{Arute2019,
  title = {Quantum Supremacy Using a Programmable Superconducting Processor},
  author = {Arute, Frank and Arya, Kunal and Babbush, Ryan and Bacon, Dave and Bardin, Joseph C. and Barends, Rami and Biswas, Rupak and Boixo, Sergio and Brandao, Fernando G. S. L. and Buell, David A. and Burkett, Brian and Chen, Yu and Chen, Zijun and Chiaro, Ben and Collins, Roberto and Courtney, William and Dunsworth, Andrew and Farhi, Edward and Foxen, Brooks and Fowler, Austin and Gidney, Craig and Giustina, Marissa and Graff, Rob and Guerin, Keith and Habegger, Steve and Harrigan, Matthew P. and Hartmann, Michael J. and Ho, Alan and Hoffmann, Markus and Huang, Trent and Humble, Travis S. and Isakov, Sergei V. and Jeffrey, Evan and Jiang, Zhang and Kafri, Dvir and Kechedzhi, Kostyantyn and Kelly, Julian and Klimov, Paul V. and Knysh, Sergey and Korotkov, Alexander and Kostritsa, Fedor and Landhuis, David and Lindmark, Mike and Lucero, Erik and Lyakh, Dmitry and Mandr{\`a}, Salvatore and McClean, Jarrod R. and McEwen, Matthew and Megrant, Anthony and Mi, Xiao and Michielsen, Kristel and Mohseni, Masoud and Mutus, Josh and Naaman, Ofer and Neeley, Matthew and Neill, Charles and Niu, Murphy Yuezhen and Ostby, Eric and Petukhov, Andre and Platt, John C. and Quintana, Chris and Rieffel, Eleanor G. and Roushan, Pedram and Rubin, Nicholas C. and Sank, Daniel and Satzinger, Kevin J. and Smelyanskiy, Vadim and Sung, Kevin J. and Trevithick, Matthew D. and Vainsencher, Amit and Villalonga, Benjamin and White, Theodore and Yao, Z. Jamie and Yeh, Ping and Zalcman, Adam and Neven, Hartmut and Martinis, John M.},
  year = {2019},
  month = oct,
  journal = {Nature},
  volume = {574},
  number = {7779},
  pages = {505--510},
  publisher = {Nature Publishing Group},
  issn = {1476-4687},
  doi = {10.1038/s41586-019-1666-5},
  urldate = {2025-05-15},
  copyright = {2019 The Author(s), under exclusive licence to Springer Nature Limited},
  langid = {english}
}

@article{Ballance2016,
  title = {High-{{Fidelity Quantum Logic Gates Using Trapped-Ion Hyperfine Qubits}}},
  author = {Ballance, C. J. and Harty, T. P. and Linke, N. M. and Sepiol, M. A. and Lucas, D. M.},
  year = {2016},
  month = aug,
  journal = {Phys. Rev. Lett.},
  volume = {117},
  number = {6},
  pages = {060504},
  publisher = {American Physical Society},
  doi = {10.1103/PhysRevLett.117.060504},
  urldate = {2024-09-18}
}

@article{Bentley2013,
  title = {Fast Gates for Ion Traps by Splitting Laser Pulses},
  author = {Bentley, C. D. B. and Carvalho, A. R. R. and Kielpinski, D. and Hope, J. J.},
  year = {2013},
  month = apr,
  journal = {New J. Phys.},
  volume = {15},
  number = {4},
  pages = {043006},
  publisher = {IOP Publishing},
  issn = {1367-2630},
  doi = {10.1088/1367-2630/15/4/043006},
  urldate = {2024-02-12},
  langid = {english}
}

@article{Bentley2016a,
  title = {Stability Thresholds and Calculation Techniques for Fast Entangling Gates on Trapped Ions},
  author = {Bentley, C. D. B. and Taylor, R. L. and Carvalho, A. R. R. and Hope, J. J.},
  year = {2016},
  month = apr,
  journal = {Phys. Rev. A},
  volume = {93},
  number = {4},
  pages = {042342},
  publisher = {American Physical Society},
  doi = {10.1103/PhysRevA.93.042342},
  urldate = {2025-05-01}
}

@article{Bentley2020,
  title = {Numeric {{Optimization}} for {{Configurable}}, {{Parallel}}, {{Error-Robust Entangling Gates}} in {{Large Ion Registers}}},
  author = {Bentley, Christopher D. B. and Ball, Harrison and Biercuk, Michael J. and Carvalho, Andre R. R. and Hush, Michael R. and Slatyer, Harry J.},
  year = {2020},
  journal = {Advanced Quantum Technologies},
  volume = {3},
  number = {11},
  pages = {2000044},
  issn = {2511-9044},
  doi = {10.1002/qute.202000044},
  urldate = {2025-05-01},
  copyright = {{\copyright} 2020 WILEY-VCH Verlag GmbH \& Co. KGaA, Weinheim},
  langid = {english}
}

@article{Bravyi2024,
  title = {High-Threshold and Low-Overhead Fault-Tolerant Quantum Memory},
  author = {Bravyi, Sergey and Cross, Andrew W. and Gambetta, Jay M. and Maslov, Dmitri and Rall, Patrick and Yoder, Theodore J.},
  year = {2024},
  month = mar,
  journal = {Nature},
  volume = {627},
  number = {8005},
  pages = {778--782},
  publisher = {Nature Publishing Group},
  issn = {1476-4687},
  doi = {10.1038/s41586-024-07107-7},
  urldate = {2025-05-15},
  copyright = {2024 The Author(s)},
  langid = {english}
}

@article{Bruzewicz2019b,
  title = {Trapped-Ion Quantum Computing: {{Progress}} and Challenges},
  shorttitle = {Trapped-Ion Quantum Computing},
  author = {Bruzewicz, Colin D. and Chiaverini, John and McConnell, Robert and Sage, Jeremy M.},
  year = {2019},
  month = may,
  journal = {Applied Physics Reviews},
  volume = {6},
  number = {2},
  pages = {021314},
  doi = {10.1063/1.5088164},
  urldate = {2019-10-02}
}

@article{Cai2023b,
  title = {Entangling Gates for Trapped-Ion Quantum Computation and Quantum Simulation},
  author = {Cai, Zhengyang and Luan, Chun -Yang and Ou, Lingfeng and Tu, Hengchao and Yin, Zihan and Zhang, Jing -Ning and Kim, Kihwan},
  year = {2023},
  month = may,
  journal = {J. Korean Phys. Soc.},
  volume = {82},
  number = {9},
  pages = {882--900},
  issn = {1976-8524},
  doi = {10.1007/s40042-023-00772-3},
  urldate = {2025-05-15},
  langid = {english}
}

@article{Campbell2010a,
  title = {Ultrafast {{Gates}} for {{Single Atomic Qubits}}},
  author = {Campbell, W. C. and Mizrahi, J. and Quraishi, Q. and Senko, C. and Hayes, D. and Hucul, D. and Matsukevich, D. N. and Maunz, P. and Monroe, C.},
  year = {2010},
  month = aug,
  journal = {Phys. Rev. Lett.},
  volume = {105},
  number = {9},
  pages = {090502},
  publisher = {American Physical Society},
  doi = {10.1103/PhysRevLett.105.090502},
  urldate = {2024-08-27}
}

@article{Cetina2022,
  title = {Control of {{Transverse Motion}} for {{Quantum Gates}} on {{Individually Addressed Atomic Qubits}}},
  author = {Cetina, M. and Egan, L.N. and Noel, C. and Goldman, M.L. and Biswas, D. and Risinger, A.R. and Zhu, D. and Monroe, C.},
  year = {2022},
  month = mar,
  journal = {PRX Quantum},
  volume = {3},
  number = {1},
  pages = {010334},
  publisher = {American Physical Society},
  doi = {10.1103/PRXQuantum.3.010334},
  urldate = {2025-05-15}
}

@article{Chen2024,
  title = {Benchmarking a Trapped-Ion Quantum Computer with 30 Qubits},
  author = {Chen, Jwo-Sy and Nielsen, Erik and Ebert, Matthew and Inlek, Volkan and Wright, Kenneth and Chaplin, Vandiver and Maksymov, Andrii and P{\'a}ez, Eduardo and Poudel, Amrit and Maunz, Peter and Gamble, John},
  year = {2024},
  month = nov,
  journal = {Quantum},
  volume = {8},
  pages = {1516},
  publisher = {Verein zur F{\"o}rderung des Open Access Publizierens in den Quantenwissenschaften},
  doi = {10.22331/q-2024-11-07-1516},
  urldate = {2025-04-16},
  langid = {british}
}

@article{Cheng2023a,
  title = {High-Quality Parallel Entangling Gates in Long Mixed-Species Ion Chains},
  author = {Cheng, Lin and Liu, Sheng-Chen and Yao, Gui-Zhong and Wang, Ying-Xiang and Peng, Liang-You and Gong, Qihuang},
  year = {2023},
  month = oct,
  journal = {Phys. Rev. A},
  volume = {108},
  number = {4},
  pages = {042603},
  publisher = {American Physical Society},
  doi = {10.1103/PhysRevA.108.042603},
  urldate = {2024-08-28}
}

@article{Cheng2024,
  title = {Crosstalk Suppression of Parallel Gates for Fault-Tolerant Quantum Computation with Trapped Ions via Optical Tweezers},
  author = {Cheng, Lin and Liu, Sheng-Chen and Peng, Liang-You and Gong, Qihuang},
  year = {2024},
  month = sep,
  journal = {Phys. Rev. Appl.},
  volume = {22},
  number = {3},
  pages = {034021},
  publisher = {American Physical Society},
  doi = {10.1103/PhysRevApplied.22.034021},
  urldate = {2025-01-10}
}

@article{Debnath2016,
  title = {Demonstration of a Small Programmable Quantum Computer with Atomic Qubits},
  author = {Debnath, S. and Linke, N. M. and Figgatt, C. and Landsman, K. A. and Wright, K. and Monroe, C.},
  year = {2016},
  month = aug,
  journal = {Nature},
  volume = {536},
  number = {7614},
  pages = {63--66},
  publisher = {Nature Publishing Group},
  issn = {1476-4687},
  doi = {10.1038/nature18648},
  urldate = {2025-04-16},
  copyright = {2016 Macmillan Publishers Limited, part of Springer Nature. All rights reserved.},
  langid = {english}
}

@article{Drmota2023a,
  title = {Robust {{Quantum Memory}} in a {{Trapped-Ion Quantum Network Node}}},
  author = {Drmota, P. and Main, D. and Nadlinger, D. P. and Nichol, B. C. and Weber, M. A. and Ainley, E. M. and Agrawal, A. and Srinivas, R. and Araneda, G. and Ballance, C. J. and Lucas, D. M.},
  year = {2023},
  month = mar,
  journal = {Phys. Rev. Lett.},
  volume = {130},
  number = {9},
  pages = {090803},
  publisher = {American Physical Society},
  doi = {10.1103/PhysRevLett.130.090803},
  urldate = {2024-04-03}
}

@article{Drmota2024,
  title = {Verifiable {{Blind Quantum Computing}} with {{Trapped Ions}} and {{Single Photons}}},
  author = {Drmota, P. and Nadlinger, D. P. and Main, D. and Nichol, B. C. and Ainley, E. M. and Leichtle, D. and Mantri, A. and Kashefi, E. and Srinivas, R. and Araneda, G. and Ballance, C. J. and Lucas, D. M.},
  year = {2024},
  month = apr,
  journal = {Phys. Rev. Lett.},
  volume = {132},
  number = {15},
  pages = {150604},
  publisher = {American Physical Society},
  doi = {10.1103/PhysRevLett.132.150604},
  urldate = {2025-05-01}
}

@article{Duan2001,
  title = {Geometric Manipulation of Trapped Ions for Quantum Computation},
  author = {Duan, L.-M.},
  year = {2001},
  month = jun,
  journal = {Science},
  volume = {292},
  number = {5522},
  pages = {1695--1697},
  issn = {00368075},
  doi = {10.1126/science.1058835},
  arxiv = {0111086 [quant-ph]},
  arxivid = {quant-ph/0111086},
  isbn = {0036-8075},
  pmid = {11387469}
}

@article{Duan2004a,
  title = {Scaling {{Ion Trap Quantum Computation}} through {{Fast Quantum Gates}}},
  author = {Duan, L.-M.},
  year = {2004},
  month = sep,
  journal = {Phys. Rev. Lett.},
  volume = {93},
  number = {10},
  pages = {100502},
  publisher = {American Physical Society},
  doi = {10.1103/PhysRevLett.93.100502},
  urldate = {2025-05-01}
}

@article{Egan2021,
  title = {Fault-Tolerant Control of an Error-Corrected Qubit},
  author = {Egan, Laird and Debroy, Dripto M. and Noel, Crystal and Risinger, Andrew and Zhu, Daiwei and Biswas, Debopriyo and Newman, Michael and Li, Muyuan and Brown, Kenneth R. and Cetina, Marko and Monroe, Christopher},
  year = {2021},
  month = oct,
  journal = {Nature},
  volume = {598},
  number = {7880},
  pages = {281--286},
  publisher = {Nature Publishing Group},
  issn = {1476-4687},
  doi = {10.1038/s41586-021-03928-y},
  urldate = {2025-05-15},
  copyright = {2021 The Author(s), under exclusive licence to Springer Nature Limited},
  langid = {english}
}

@article{Figgatt2019,
  title = {Parallel Entangling Operations on a Universal Ion-Trap Quantum Computer},
  author = {Figgatt, C. and Ostrander, A. and Linke, N. M. and Landsman, K. A. and Zhu, D. and Maslov, D. and Monroe, C.},
  year = {2019},
  month = aug,
  journal = {Nature},
  volume = {572},
  number = {7769},
  pages = {368--372},
  publisher = {Nature Publishing Group},
  issn = {1476-4687},
  doi = {10.1038/s41586-019-1427-5},
  urldate = {2024-08-27},
  copyright = {2019 The Author(s), under exclusive licence to Springer Nature Limited},
  langid = {english}
}

@article{Gale2020,
  title = {Optimized Fast Gates for Quantum Computing with Trapped Ions},
  author = {Gale, {\relax EPG} and Mehdi, Z and Oberg, {\relax LM} and {AK Ratcliffe} and {SA Haine} and {JJ Hope}},
  year = {2020},
  month = may,
  journal = {Phys. Rev. A},
  volume = {101},
  number = {5},
  pages = {052328},
  publisher = {American Physical Society},
  doi = {10.1103/PhysRevA.101.052328}
}

@article{Gale2020a,
  title = {Optimized Fast Gates for Quantum Computing with Trapped Ions},
  author = {Gale, Evan P. G. and Mehdi, Zain and Oberg, Lachlan M. and Ratcliffe, Alexander K. and Haine, Simon A. and Hope, Joseph J.},
  year = {2020},
  month = may,
  journal = {Phys. Rev. A},
  volume = {101},
  number = {5},
  pages = {052328},
  publisher = {American Physical Society},
  doi = {10.1103/PhysRevA.101.052328},
  urldate = {2023-12-10}
}

@article{Garcia-Ripoll2003,
  title = {Speed Optimized Two-Qubit Gates with Laser Coherent Control Techniques for Ion Trap Quantum Computing},
  author = {{Garc{\'i}a-Ripoll}, J. J. and Zoller, P. and Cirac, J. I.},
  year = {2003},
  month = oct,
  journal = {Physical Review Letters},
  volume = {91},
  number = {15},
  pages = {157901},
  publisher = {APS},
  issn = {0031-9007},
  doi = {10.1103/PhysRevLett.91.157901},
  arxiv = {0306006 [quant-ph]},
  arxivid = {quant-ph/0306006},
  isbn = {0031-9007},
  pmid = {14611499}
}

@article{Garcia-Ripoll2005,
  title = {Coherent Control of Trapped Ions Using Off-Resonant Lasers},
  author = {{Garcia-Ripoll}, J. J. and Zoller, P. and Cirac, J. I.},
  year = {2005},
  month = jun,
  journal = {Phys. Rev. A},
  volume = {71},
  number = {6},
  eprint = {quant-ph/0411103},
  pages = {062309},
  issn = {1050-2947, 1094-1622},
  doi = {10.1103/PhysRevA.71.062309},
  urldate = {2019-06-01},
  archiveprefix = {arXiv},
  langid = {english}
}

@article{Graham2022,
  title = {Multi-Qubit Entanglement and Algorithms on a Neutral-Atom Quantum Computer},
  author = {Graham, T. M. and Song, Y. and Scott, J. and Poole, C. and Phuttitarn, L. and Jooya, K. and Eichler, P. and Jiang, X. and Marra, A. and Grinkemeyer, B. and Kwon, M. and Ebert, M. and Cherek, J. and Lichtman, M. T. and Gillette, M. and Gilbert, J. and Bowman, D. and Ballance, T. and Campbell, C. and Dahl, E. D. and Crawford, O. and Blunt, N. S. and Rogers, B. and Noel, T. and Saffman, M.},
  year = {2022},
  month = apr,
  journal = {Nature},
  volume = {604},
  number = {7906},
  pages = {457--462},
  publisher = {Nature Publishing Group},
  issn = {1476-4687},
  doi = {10.1038/s41586-022-04603-6},
  urldate = {2025-05-15},
  copyright = {2022 The Author(s), under exclusive licence to Springer Nature Limited},
  langid = {english}
}

@article{Grzesiak2020,
  title = {Efficient Arbitrary Simultaneously Entangling Gates on a Trapped-Ion Quantum Computer},
  author = {Grzesiak, Nikodem and Bl{\"u}mel, Reinhold and Wright, Kenneth and Beck, Kristin M. and Pisenti, Neal C. and Li, Ming and Chaplin, Vandiver and Amini, Jason M. and Debnath, Shantanu and Chen, Jwo-Sy and Nam, Yunseong},
  year = {2020},
  month = jun,
  journal = {Nat Commun},
  volume = {11},
  number = {1},
  pages = {2963},
  publisher = {Nature Publishing Group},
  issn = {2041-1723},
  doi = {10.1038/s41467-020-16790-9},
  urldate = {2024-08-27},
  copyright = {2020 The Author(s)},
  langid = {english}
}

@article{Grzesiak2020a,
  title = {Efficient Arbitrary Simultaneously Entangling Gates on a Trapped-Ion Quantum Computer},
  author = {Grzesiak, Nikodem and Bl{\"u}mel, Reinhold and Wright, Kenneth and Beck, Kristin M. and Pisenti, Neal C. and Li, Ming and Chaplin, Vandiver and Amini, Jason M. and Debnath, Shantanu and Chen, Jwo-Sy and Nam, Yunseong},
  year = {2020},
  month = jun,
  journal = {Nat Commun},
  volume = {11},
  number = {1},
  pages = {2963},
  publisher = {Nature Publishing Group},
  issn = {2041-1723},
  doi = {10.1038/s41467-020-16790-9},
  urldate = {2025-01-10},
  copyright = {2020 The Author(s)},
  langid = {english}
}

@article{Guo2022b,
  title = {Picosecond Ion-Qubit Manipulation and Spin-Phonon Entanglement with Resonant Laser Pulses},
  author = {Guo, W.-X. and Wu, Y.-K. and Huang, Y.-Y. and Feng, L. and Huang, C.-X. and Yang, H.-X. and Ma, J.-Y. and Yao, L. and Zhou, Z.-C. and Duan, L.-M.},
  year = {2022},
  month = aug,
  journal = {Phys. Rev. A},
  volume = {106},
  number = {2},
  pages = {022608},
  publisher = {American Physical Society},
  doi = {10.1103/PhysRevA.106.022608},
  urldate = {2024-08-27}
}

@article{Harty2014,
  title = {High-{{Fidelity Preparation}}, {{Gates}}, {{Memory}}, and {{Readout}} of a {{Trapped-Ion Quantum Bit}}},
  author = {Harty, T. P. and Allcock, D. T. C. and Ballance, C. J. and Guidoni, L. and Janacek, H. A. and Linke, N. M. and Stacey, D. N. and Lucas, D. M.},
  year = {2014},
  month = nov,
  journal = {Phys. Rev. Lett.},
  volume = {113},
  number = {22},
  pages = {220501},
  publisher = {American Physical Society},
  doi = {10.1103/PhysRevLett.113.220501},
  urldate = {2025-05-15}
}

@article{Heinrich2019a,
  title = {Ultrafast Coherent Excitation of a {{40Ca}}+ Ion},
  author = {Heinrich, D. and Guggemos, M. and {Guevara-Bertsch}, M. and Hussain, M. I. and Roos, C. F. and Blatt, R.},
  year = {2019},
  month = jul,
  journal = {New J. Phys.},
  volume = {21},
  number = {7},
  pages = {073017},
  publisher = {IOP Publishing},
  issn = {1367-2630},
  doi = {10.1088/1367-2630/ab2a7e},
  urldate = {2024-08-27},
  langid = {english}
}

@article{Hempel2018a,
  title = {Quantum {{Chemistry Calculations}} on a {{Trapped-Ion Quantum Simulator}}},
  author = {Hempel, Cornelius and Maier, Christine and Romero, Jonathan and McClean, Jarrod and Monz, Thomas and Shen, Heng and Jurcevic, Petar and Lanyon, Ben P. and Love, Peter and Babbush, Ryan and {Aspuru-Guzik}, Al{\'a}n and Blatt, Rainer and Roos, Christian F.},
  year = {2018},
  month = jul,
  journal = {Phys. Rev. X},
  volume = {8},
  number = {3},
  pages = {031022},
  publisher = {American Physical Society},
  doi = {10.1103/PhysRevX.8.031022},
  urldate = {2025-05-15}
}

@article{Home2011,
  title = {Normal Modes of Trapped Ions in the Presence of Anharmonic Trap Potentials},
  author = {Home, J P and Hanneke, D and Jost, J D and Leibfried, D and Wineland, D J},
  year = {2011},
  month = jul,
  journal = {New J. Phys.},
  volume = {13},
  number = {7},
  pages = {073026},
  issn = {1367-2630},
  doi = {10.1088/1367-2630/13/7/073026},
  urldate = {2024-04-23}
}

@article{Hou2024,
  title = {Individually Addressed Entangling Gates in a Two-Dimensional Ion Crystal},
  author = {Hou, Y.-H. and Yi, Y.-J. and Wu, Y.-K. and Chen, Y.-Y. and Zhang, L. and Wang, Y. and Xu, Y.-L. and Zhang, C. and Mei, Q.-X. and Yang, H.-X. and Ma, J.-Y. and Guo, S.-A. and Ye, J. and Qi, B.-X. and Zhou, Z.-C. and Hou, P.-Y. and Duan, L.-M.},
  year = {2024},
  month = nov,
  journal = {Nat Commun},
  volume = {15},
  number = {1},
  pages = {9710},
  publisher = {Nature Publishing Group},
  issn = {2041-1723},
  doi = {10.1038/s41467-024-53405-z},
  urldate = {2025-01-10},
  copyright = {2024 The Author(s)},
  langid = {english}
}

@article{James1998a,
  title = {Quantum Dynamics of Cold Trapped Ions, with Application to Quantum Computation},
  author = {James, Daniel F. V.},
  year = {1998},
  month = feb,
  journal = {Applied Physics B: Lasers and Optics},
  volume = {66},
  number = {2},
  eprint = {quant-ph/9702053},
  pages = {181--190},
  issn = {0946-2171, 1432-0649},
  doi = {10.1007/s003400050373},
  urldate = {2019-02-23},
  archiveprefix = {arXiv}
}

@article{Johnson2015,
  title = {Sensing {{Atomic Motion}} from the {{Zero Point}} to {{Room Temperature}} with {{Ultrafast Atom Interferometry}}},
  author = {Johnson, K. G. and Neyenhuis, B. and Mizrahi, J. and {Wong-Campos}, J. D. and Monroe, C.},
  year = {2015},
  month = nov,
  journal = {Phys. Rev. Lett.},
  volume = {115},
  number = {21},
  pages = {213001},
  publisher = {American Physical Society},
  doi = {10.1103/PhysRevLett.115.213001},
  urldate = {2025-05-15}
}

@article{Kaushal2020a,
  title = {Shuttling-Based Trapped-Ion Quantum Information Processing},
  author = {Kaushal, V. and Lekitsch, B. and Stahl, A. and Hilder, J. and Pijn, D. and Schmiegelow, C. and Bermudez, A. and M{\"u}ller, M. and {Schmidt-Kaler}, F. and Poschinger, U.},
  year = {2020},
  month = mar,
  journal = {AVS Quantum Science},
  volume = {2},
  number = {1},
  pages = {014101},
  issn = {2639-0213},
  doi = {10.1116/1.5126186},
  urldate = {2025-05-15}
}

@article{Kielpinski2002,
  title = {Architecture for a Large-Scale Ion-Trap Quantum Computer},
  author = {Kielpinski, D. and Monroe, C. and Wineland, D. J.},
  year = {2002},
  month = jun,
  journal = {Nature},
  volume = {417},
  number = {6890},
  pages = {709--711},
  publisher = {Nature Publishing Group},
  issn = {1476-4687},
  doi = {10.1038/nature00784},
  urldate = {2025-05-15},
  copyright = {2002 Macmillan Magazines Ltd.},
  langid = {english}
}

@article{Landsman2019,
  title = {Two-Qubit Entangling Gates within Arbitrarily Long Chains of Trapped Ions},
  author = {Landsman, K. A. and Wu, Y. and Leung, P. H. and Zhu, D. and Linke, N. M. and Brown, K. R. and Duan, L. and Monroe, C.},
  year = {2019},
  month = aug,
  journal = {Phys. Rev. A},
  volume = {100},
  number = {2},
  pages = {022332},
  publisher = {American Physical Society},
  doi = {10.1103/PhysRevA.100.022332},
  urldate = {2025-01-10}
}

@article{Langer2005,
  title = {Long-{{Lived Qubit Memory Using Atomic Ions}}},
  author = {Langer, C. and Ozeri, R. and Jost, J. D. and Chiaverini, J. and DeMarco, B. and {Ben-Kish}, A. and Blakestad, R. B. and Britton, J. and Hume, D. B. and Itano, W. M. and Leibfried, D. and Reichle, R. and Rosenband, T. and Schaetz, T. and Schmidt, P. O. and Wineland, D. J.},
  year = {2005},
  month = aug,
  journal = {Phys. Rev. Lett.},
  volume = {95},
  number = {6},
  pages = {060502},
  publisher = {American Physical Society},
  doi = {10.1103/PhysRevLett.95.060502},
  urldate = {2025-05-15}
}

@article{Lee2005,
  title = {Phase Control of Trapped Ion Quantum Gates},
  author = {Lee, P J and Brickman, K-A and Deslauriers, L and Haljan, P C and Duan, L-M and Monroe, C},
  year = {2005},
  month = oct,
  journal = {J. Opt. B: Quantum Semiclass. Opt.},
  volume = {7},
  number = {10},
  pages = {S371-S383},
  issn = {1464-4266, 1741-3575},
  doi = {10.1088/1464-4266/7/10/025},
  urldate = {2020-02-03},
  langid = {english}
}

@article{Leu_2023,
  title = {Fast, High-Fidelity Addressed Single-Qubit Gates Using Efficient Composite Pulse Sequences},
  author = {Leu, A. D. and Gely, M. F. and Weber, M. A. and Smith, M. C. and Nadlinger, D. P. and Lucas, D. M.},
  year = {2023},
  month = sep,
  journal = {Phys. Rev. Lett.},
  volume = {131},
  number = {12},
  pages = {120601},
  publisher = {American Physical Society},
  doi = {10.1103/PhysRevLett.131.120601}
}

@article{Leung2018,
  title = {Entangling an Arbitrary Pair of Qubits in a Long Ion Crystal},
  author = {Leung, Pak Hong and Brown, Kenneth R.},
  year = {2018},
  month = sep,
  journal = {Phys. Rev. A},
  volume = {98},
  number = {3},
  pages = {032318},
  publisher = {American Physical Society},
  doi = {10.1103/PhysRevA.98.032318},
  urldate = {2025-01-10}
}

@article{Li2025,
  title = {Realizing Two-Qubit Gates through Mode Engineering on a Trapped-Ion Quantum Computer},
  author = {Li, Ming and Nguyen, Nhung H. and Green, Alaina M. and Amini, Jason and Linke, Norbert M. and Nam, Yunseong},
  year = {2025},
  month = feb,
  journal = {Phys. Rev. A},
  volume = {111},
  number = {2},
  pages = {022622},
  publisher = {American Physical Society},
  doi = {10.1103/PhysRevA.111.022622},
  urldate = {2025-05-15}
}

@article{Linke2017,
  title = {Experimental Comparison of Two Quantum Computing Architectures},
  author = {Linke, Norbert M. and Maslov, Dmitri and Roetteler, Martin and Debnath, Shantanu and Figgatt, Caroline and Landsman, Kevin A. and Wright, Kenneth and Monroe, Christopher},
  year = {2017},
  month = mar,
  journal = {Proceedings of the National Academy of Sciences},
  volume = {114},
  number = {13},
  pages = {3305--3310},
  publisher = {Proceedings of the National Academy of Sciences},
  doi = {10.1073/pnas.1618020114},
  urldate = {2025-05-15}
}

@article{Liu2023,
  title = {Efficient Numerical Approach to High-Fidelity Phase-Modulated Gates in Long Chains of Trapped Ions},
  author = {Liu, Sheng-Chen and Cheng, Lin and Yao, Gui-Zhong and Wang, Ying-Xiang and Peng, Liang-You},
  year = {2023},
  month = mar,
  journal = {Phys. Rev. E},
  volume = {107},
  number = {3},
  pages = {035304},
  publisher = {American Physical Society},
  doi = {10.1103/PhysRevE.107.035304},
  urldate = {2025-01-10}
}

@article{Liu2025,
  title = {Certified Randomness Using a Trapped-Ion Quantum Processor},
  author = {Liu, Minzhao and Shaydulin, Ruslan and Niroula, Pradeep and DeCross, Matthew and Hung, Shih-Han and Kon, Wen Yu and {Cervero-Mart{\'i}n}, Enrique and Chakraborty, Kaushik and Amer, Omar and Aaronson, Scott and Acharya, Atithi and Alexeev, Yuri and Berg, K. Jordan and Chakrabarti, Shouvanik and Curchod, Florian J. and Dreiling, Joan M. and Erickson, Neal and Foltz, Cameron and {Foss-Feig}, Michael and Hayes, David and Humble, Travis S. and Kumar, Niraj and Larson, Jeffrey and Lykov, Danylo and Mills, Michael and Moses, Steven A. and Neyenhuis, Brian and Eloul, Shaltiel and Siegfried, Peter and Walker, James and Lim, Charles and Pistoia, Marco},
  year = {2025},
  month = apr,
  journal = {Nature},
  volume = {640},
  number = {8058},
  pages = {343--348},
  publisher = {Nature Publishing Group},
  issn = {1476-4687},
  doi = {10.1038/s41586-025-08737-1},
  urldate = {2025-05-15},
  copyright = {2025 The Author(s)},
  langid = {english}
}

@misc{Main2024,
  title = {Distributed {{Quantum Computing}} across an {{Optical Network Link}}},
  author = {Main, D. and Drmota, P. and Nadlinger, D. P. and Ainley, E. M. and Agrawal, A. and Nichol, B. C. and Srinivas, R. and Araneda, G. and Lucas, D. M.},
  year = {2024},
  month = jun,
  journal = {arXiv.org},
  urldate = {2024-09-06},
  howpublished = {https://arxiv.org/abs/2407.00835v1},
  langid = {english}
}

@article{Main2025,
  title = {Distributed Quantum Computing across an Optical Network Link},
  author = {Main, D. and Drmota, P. and Nadlinger, D. P. and Ainley, E. M. and Agrawal, A. and Nichol, B. C. and Srinivas, R. and Araneda, G. and Lucas, D. M.},
  year = {2025},
  month = feb,
  journal = {Nature},
  volume = {638},
  number = {8050},
  pages = {383--388},
  publisher = {Nature Publishing Group},
  issn = {1476-4687},
  doi = {10.1038/s41586-024-08404-x},
  urldate = {2025-05-01},
  copyright = {2025 The Author(s)},
  langid = {english}
}

@misc{Mayer2024,
  title = {Benchmarking Logical Three-Qubit Quantum {{Fourier}} Transform Encoded in the {{Steane}} Code on a Trapped-Ion Quantum Computer},
  author = {Mayer, Karl and {Ryan-Anderson}, Ciar{\'a}n and Brown, Natalie and {Durso-Sabina}, Elijah and Baldwin, Charles H. and Hayes, David and Dreiling, Joan M. and Foltz, Cameron and Gaebler, John P. and Gatterman, Thomas M. and Gerber, Justin A. and Gilmore, Kevin and Gresh, Dan and Hewitt, Nathan and Horst, Chandler V. and Johansen, Jacob and Mengle, Tanner and Mills, Michael and Moses, Steven A. and Siegfried, Peter E. and Neyenhuis, Brian and Pino, Juan and Stutz, Russell},
  year = {2024},
  month = apr,
  number = {arXiv:2404.08616},
  eprint = {2404.08616},
  primaryclass = {quant-ph},
  publisher = {arXiv},
  doi = {10.48550/arXiv.2404.08616},
  urldate = {2025-05-15},
  archiveprefix = {arXiv}
}

@article{Mehdi2020b:2D,
  title = {Scalable Quantum Computation with Fast Gates in Two-Dimensional Microtrap Arrays of Trapped Ions},
  author = {Mehdi, Z and {AK Ratcliffe} and {JJ Hope}},
  year = {2020},
  month = jul,
  journal = {Phys. Rev. A},
  volume = {102},
  number = {1},
  pages = {012618},
  publisher = {American Physical Society},
  doi = {10.1103/PhysRevA.102.012618}
}

@article{Mehdi2021,
  title = {Fast Entangling Gates in Long Ion Chains},
  author = {{Z Mehdi} and {AK Ratcliffe} and {JJ Hope}},
  year = {2021},
  month = jan,
  journal = {Phys. Rev. Res.},
  volume = {3},
  number = {1},
  pages = {013026},
  publisher = {American Physical Society},
  doi = {10.1103/PhysRevResearch.3.013026}
}

@article{Mehdi2021e,
  title = {Fast Entangling Gates in Long Ion Chains},
  author = {Mehdi, Zain and Ratcliffe, Alexander K. and Hope, Joseph J.},
  year = {2021},
  month = jan,
  journal = {Phys. Rev. Research},
  volume = {3},
  number = {1},
  pages = {013026},
  issn = {2643-1564},
  doi = {10.1103/PhysRevResearch.3.013026},
  urldate = {2024-04-23},
  langid = {english}
}

@misc{Mehdi2025,
  title = {Fast Mixed-Species Quantum Logic Gates for Trapped-Ion Quantum Networks},
  author = {Mehdi, Zain and Vaidya, Varun D. and {Savill-Brown}, Isabelle and Grosser, Phoebe and Ratcliffe, Alexander K. and Liu, Haonan and Haine, Simon A. and Hope, Joseph J. and Viteri, C. Ricardo},
  year = {2025},
  month = mar,
  number = {arXiv:2412.07185},
  eprint = {2412.07185},
  primaryclass = {quant-ph},
  publisher = {arXiv},
  doi = {10.48550/arXiv.2412.07185},
  urldate = {2025-05-01},
  archiveprefix = {arXiv}
}

@article{Mizrahi2013a,
  title = {Ultrafast {{Spin-Motion Entanglement}} and {{Interferometry}} with a {{Single Atom}}},
  author = {Mizrahi, J. and Senko, C. and Neyenhuis, B. and Johnson, K. G. and Campbell, W. C. and Conover, C. W. S. and Monroe, C.},
  year = {2013},
  month = may,
  journal = {Phys. Rev. Lett.},
  volume = {110},
  number = {20},
  eprint = {1201.6597},
  primaryclass = {physics, physics:quant-ph},
  pages = {203001},
  issn = {0031-9007, 1079-7114},
  doi = {10.1103/PhysRevLett.110.203001},
  urldate = {2024-02-13},
  archiveprefix = {arXiv},
  langid = {english}
}

@article{Mizrahi2013b,
  title = {Ultrafast {{Spin-Motion Entanglement}} and {{Interferometry}} with a {{Single Atom}}},
  author = {Mizrahi, J. and Senko, C. and Neyenhuis, B. and Johnson, K. G. and Campbell, W. C. and Conover, C. W. S. and Monroe, C.},
  year = {2013},
  month = may,
  journal = {Phys. Rev. Lett.},
  volume = {110},
  number = {20},
  pages = {203001},
  publisher = {American Physical Society},
  doi = {10.1103/PhysRevLett.110.203001},
  urldate = {2024-08-27}
}

\end{document}